\documentclass[aps,prd,eqsecnum,twocolumn,showpacs,amsmath]{revtex4}
\usepackage[dvips]{color,graphicx}
\usepackage{amsfonts,amssymb,theorem,soul,xcolor,ulem}
\newcommand{\qed}{\hbox{\rule[-2pt]{6pt}{6pt}}}
\newcommand{\D}{{\rm d}}

{\theorembodyfont{\upshape}
\newtheorem{Prop}{Proposition}}
{\theorembodyfont{\upshape}
}
{\theorembodyfont{\upshape}
\newtheorem{lm}{Lemma}}
{\theorembodyfont{\upshape}
}
{\theorembodyfont{\upshape}
}
{\theorembodyfont{\upshape}
}

\newcommand{\dalm}{\kern1pt\vbox{\hrule height 0.9pt\hbox{\vrule width
0.9pt\hskip 2.5pt\vbox{\vskip 5.5pt}\hskip 3pt\vrule width 0.3pt}\hrule height
0.3pt}\kern1pt}

\begin{document}

\title{
Exact plane symmetric black bounce with a perfect-fluid exterior \\ obeying a linear equation of state
}

\author{Hideki Maeda${}^{a}$}
\email{h-maeda@hgu.jp}
\author{Cristi{\'a}n Mart\'{\i}nez${}^{b,c}$}
\email{cristian.martinez@uss.cl}


\affiliation{
${}^a$ Department of Electronics and Information Engineering, Hokkai-Gakuen University, Sapporo 062-8605, Japan}
\affiliation{ 
${}^b$ Centro de Estudios Cient\'{\i}ficos (CECs), Avenida Arturo Prat 514, Valdivia, Chile}
\affiliation{
${}^c$ Facultad de Ingenier\'{\i}a, Universidad San Sebasti\'an, General Lagos 1163, Valdivia, Chile}

\date{\today}

\begin{abstract}
We investigate an exact two-parameter family of plane symmetric solutions admitting a hypersurface-orthogonal Killing vector in general relativity with a perfect fluid obeying a linear equation of state $p=\chi\rho$ in $n(\ge 4)$ dimensions, obtained by Gamboa in 2012.
The Gamboa solution is identical to the topological Schwarzschild-Tangherlini-(anti-)de~Sitter $\Lambda$-vacuum solution for $\chi=-1$ and admits a nondegenerate Killing horizon only for $\chi=-1$ and $\chi\in[-1/3,0)$.
We identify all possible regular attachments of two Gamboa solutions for $\chi\in[-1/3,0)$ at the Killing horizon without a lightlike thin shell, where $\chi$ may have different values on each side of the horizon.
We also present the maximal extension of the static and asymptotically topological Schwarzschild-Tangherlini Gamboa solution, realized only for $\chi\in(-(n-3)/(3n-5),0)$, under the assumption that the value of $\chi$ is unchanged in the extended dynamical region beyond the horizon.
The maximally extended spacetime describes either (i) a globally regular black bounce whose Killing horizon coincides with a bounce null hypersurface or (ii) a black hole with a spacelike curvature singularity inside the horizon.
The matter field inside the horizon is not a perfect fluid but rather an anisotropic fluid that can be interpreted as a spacelike (tachyonic) perfect fluid.
A fine-tuning of the parameters is unnecessary for the black bounce, but the null energy condition is violated everywhere except on the horizon.
In the black-bounce (black-hole) case, the metric in the regular coordinate system is $C^\infty$ only for $\chi=-1/(1+2N)$ with odd (even) $N$ satisfying $N>(n-1)/(n-3)$, and if one of the parameters in the extended region is fine-tuned.
\end{abstract}



\maketitle


\section{Introduction}

The interior structure of a realistic black hole is a highly nontrivial problem.
Although curvature singularities exist inside the Schwarzschild and Kerr vacuum black holes in general relativity, it is widely believed that the quantum effects of gravity are so dominant near the singularity that singularities do not exist as a configuration of spacetime in full quantum gravity.
Since the theory of quantum gravity is still incomplete at present, many models of singularity-free black holes have been proposed based on the belief that such {\it nonsingular black holes} are realized even in the classical theory as a low-energy limit of quantum gravity.

Among various types of nonsingular black holes, the type with a regular center has been studied in the most detail thus far. (See Sec.~2.2 in Ref.~\cite{Maeda:2021jdc} for a review.) 
In Ref.~\cite{Maeda:2021jdc}, one of the authors proposed seven criteria to single out physically reasonable nonsingular black-hole models and studied the models of Bardeen~\cite{Bardeen1968}, Hayward~\cite{Hayward2006}, Dymnikova~\cite{Dymnikova:2004zc}, and Fan and Wang~\cite{Fan:2016hvf}.
Although this type of nonsingular black hole with a regular center can be an exact solution to the Einstein equations with a nonlinear electromagnetic field~\cite{AyonBeato:2000zs}, an integration constant must be fine-tuned to remove the singularity at the center~\cite{Chinaglia:2017uqd}. (See also Appendix~A in Ref.~\cite{Maeda:2021jdc}.) 
Therefore, such nonsingular black holes are not generic configurations of black holes in that system, which does not meet one of the criteria proposed in Ref.~\cite{Maeda:2021jdc}.
Nevertheless, a modified theory of gravity is known in which electrically charged spherically symmetric black holes are generally singularity-free~\cite{Cano:2020ezi}.
Moreover, it was shown that such nonsingular black holes can be generic even in pure gravity~\cite{Bueno:2024dgm}.

On the other hand, besides the regular-center type, there is an interesting type of nonsingular black hole in which a singularity is avoided by the big bounce which occurs on a spacelike or null hypersurface in the dynamical region inside the horizon.
Such a black hole is referred to as a {\it black universe} by Bronnikov, Melnikov, and Dehnen~\cite{Bronnikov:2006fu}.
However, the catchy name {\it black bounce} introduced later by Simpson and Visser~\cite{Simpson:2018tsi} for such black holes is more popular today.
The black bounce may be more promising than the regular-center type in that it is not necessarily accompanied by an inner horizon, which is inevitable in the regular-center type and could suffer from the mass inflation instability~\cite{Poisson:1989zz,Poisson:1990eh,Ori:1991zz}.
(See Refs.~\cite{Rubio:2021obb,Bonanno:2020fgp,DiFilippo:2022qkl,Carballo-Rubio:2022kad,Carballo-Rubio:2024dca} for recent studies of the inner-horizon instability of the regular-center type.)

A black bounce has been realized as solutions with a variety of matter fields: a minimally coupled ghost scalar field with potential~\cite{Bronnikov:2005gm} and with a Maxwell field~\cite{Bolokhov:2012kn}, a nonminimally coupled scalar field with potential~\cite{Bronnikov:2022gjq}, a conformally coupled scalar field with potential and a Maxwell field~\cite{Barrientos:2022avi}, a nonlinear electromagnetic field and a minimally coupled scalar field with potential~\cite{Canate:2022gpy,Bronnikov:2022bud,Rodrigues:2023vtm,Bronnikov:2023aya,Alencar:2024yvh}, a phantom scalar field that is nonminimally coupled to the Maxwell field~\cite{Huang:2019arj,Yang:2021diz}, and $k$-essence theory, namely, a scalar field with a noncanonical kinetic term~\cite{Pereira:2023lck,Pereira:2024gsl,Pereira:2024rtv}. 
Black-bounce configurations that generalize the Simpson-Visser spacetime can be found in Refs.~\cite{Lobo:2020ffi, Franzin:2021vnj,Bronnikov:2021uta,Lessa:2024erf}. Apart from general relativity, black-bounce solutions have been found in modified theories of gravity~\cite{Junior:2022zxo,Fabris:2023opv,Junior:2024vrv,Junior:2024cbb} and in lower dimensions~\cite{Furtado:2022tnb}.
More studies concerning black bounces in the context of astrophysics examine quasinormal modes~\cite{Churilova:2019cyt,Franzin:2022iai,Wu:2022eiv}, shadows~\cite{Guerrero:2021ues,Guo:2021wid,Lima:2021las}, and gravitational lensing~\cite{Nascimento:2020ime,Tsukamoto:2020bjm,Islam:2021ful,Tsukamoto:2021caq,Cheng:2021hoc}.

In this paper, we present an exact black-bounce solution in general relativity in $n(\ge 4)$ dimensions with one of the simplest matter fields, a perfect fluid obeying a linear equation of state, instead of more complex matter fields as in most of the previously cited references. Such a simple black bounce can be achieved via a careful analysis of the causal and global structures of a static solution with planar symmetry found by Gamboa in Ref.~\cite{GamboaSaravi} 

The organization of this paper is as follows.
In Sec.~\ref{sec:pre}, we present a new form of the Gamboa solution~\cite{GamboaSaravi} and review several physical and mathematical concepts for the subsequent sections.
In Sec.~\ref{sec:extension}, we identify all possible regular attachments of two Gamboa solutions at a nondegenerate Killing horizon and derive the conditions for the $C^\infty$ attachment.
In Sec.~\ref{sec:main}, we clarify the maximally extended spacetimes of the static Gamboa solution for $\chi\in(-(n-3)/(3n-5),0)$ under the assumption that the value of $\chi$ is unchanged in the extended dynamical region beyond the horizon.
Our main results are summarized in the final section.
In Appendix~\ref{app:General-sol}, we present another derivation of the Gamboa solution generalizing the original one with an $(n-2)$-dimensional Ricci-flat base manifold.
In Appendix~\ref{app:bifurcation}, we show the regularity of the bifurcation $(n-2)$-surface in the solution.
In Appendix~\ref{app:geodesic}, we derive an explicit form of the matter field confined on the Killing horizon in the solution for $\chi=-1/3$.

Our conventions for curvature tensors are $[\nabla _\rho ,\nabla_\sigma]V^\mu ={{R}^\mu }_{\nu\rho\sigma}V^\nu$ and ${R}_{\mu \nu }={{R}^\rho }_{\mu \rho \nu }$, where Greek indices run over all spacetime indices.
The signature of the Minkowski spacetime is $(-,+,+,\cdots,+)$.
We adopt the units such that $c=1$ and $\kappa_n:=8\pi G_n$, where $G_n$ is the $n$-dimensional gravitational constant.

\section{Exact perfect-fluid solution for $p=\chi\rho$ ($\chi\ne 0,1$) with plane symmetry}
\label{sec:pre}

In this paper, we study an exact static solution to the following Einstein equations with a perfect fluid obeying $p=\chi\rho$ in $n(\ge 4)$ dimensions:
\begin{align}
\label{EFE-0}
\begin{aligned}
&G_{\mu\nu}=\kappa_nT_{\mu\nu},\\
&{T}_{\mu\nu}=(\rho+p) u_\mu u_\nu+pg_{\mu\nu}.
\end{aligned}
\end{align}
Here $\rho$ and $p$ are the energy density and pressure of a perfect fluid, respectively, and $u^\mu$ is the normalized $n$-velocity of the fluid element satisfying $u_\mu u^\mu=-1$.

The general static solution with planar symmetry in this system was obtained by Gamboa~\cite{GamboaSaravi}. 
In Appendix~\ref{app:General-sol}, using a different coordinate system, we present a simple derivation of the Gamboa solution and its generalization by considering an arbitrary Ricci-flat base manifold instead of just $\mathbb{R}^{n-2}$. However, in this work, this generalization is not considered and the base manifold is chosen as an $(n-2)$-dimensional flat space to preserve planar symmetry, whose line element is denoted by $\D l_{n-2}^2$.

\subsection{New form of the Gamboa solution for $\chi\ne 0,1$ with $A_0\ne 0$}

We are interested in the general case of the Gamboa solution for $\chi\ne 0,1$ since there is no static solution for $\chi=0$ and the solution does not admit a Killing horizon for $\chi=1$.
The forms of the Gamboa solution are different for $\chi=\chi_0$ and $\chi\ne \chi_0$, where
\begin{align}
\label{def-chi0}
\chi_0:=-\frac{n-3}{3n-5}\left(>-\frac13\right).
\end{align} 
(See Eq.~\eqref{hApend}.) By the following coordinate transformation and reparametrizations,
\begin{align}
\label{ansatz00-02}
\begin{aligned} 
&T\to A_0^{(1+\beta)/(2+\beta)} B_0^{-1/2}T,\\
&M:=-A_0^{(3+2\beta)/(2+\beta)}/(2B_0),\\
&h_1:=-A_1/A_0,
\end{aligned} 
\end{align} 
we write the Gamboa solution shown in Appendix~\ref{app:General-sol} for $\chi\ne 0,1$ with $A_0\ne 0$ in the following form: 
\begin{align}
\label{ansatz00-2}
\begin{aligned} 
&\D s^2=-A\D T^2+B\D r^2+r^2\D l_{n-2}^2,\\
&A(r)=-\frac{2M}{r^{n-3}}\Pi^{1/(2+\beta)},\\
&B(r)=-\frac{r^{n-3}}{2M}\Pi^{-(3+2\beta)/(2+\beta)},\\
&\rho=\frac{p}{\chi}=\frac{\alpha M}{r^{(n-3)(1+\beta)}} \Pi^{(1+\beta)/(2+\beta)},
\end{aligned} 
\end{align} 
where $\Pi(r)$ is defined by 
\begin{align}
\label{def-Pi}
&\Pi(r):=\left\{
\begin{array}{ll}
1-h_1/r^{(n-3)\beta-2} & [\chi\ne 0,\chi_0,1],\\
1-h_1\ln|r| & [\chi=\chi_0].
\end{array}
\right.
\end{align} 
The Gamboa solution (\ref{ansatz00-2}) is parametrized by $M(\ne 0)$ and $h_1$, and we have introduced constants $\beta$, $\zeta$, and $\alpha$ such that 
\begin{align}
&\beta:=-\frac{1+3\chi}{2\chi}~~\Leftrightarrow~~\chi=-\frac{1}{3+2\beta},\label{def-beta}\\
&\zeta:=(3n-5)\chi+(n-3)=\frac{2(n-3)\beta-4}{3+2\beta},\label{def-zeta}\\
&\alpha:=\left\{
\begin{array}{ll}
-(n-2) \zeta h_1/[\kappa_n \chi (1-\chi)] & [\chi\ne 0,\chi_0,1],\\
-(3n-5)h_1/(2\kappa_n) & [\chi=\chi_0],
\end{array}
\right.\label{def-alpha}
\end{align} 
which show $\beta=2/(n-3)$ for $\chi=\chi_0$.

For $\chi=-1$ ($\beta=-1$), the Gamboa solution (\ref{ansatz00-2}) is identical to the topological Schwarzschild-Tangherlini-(anti-)de~Sitter $\Lambda$-vacuum solution given by 
\begin{align}
\label{ansatz00-2-Lambda}
\begin{aligned} 
&\D s^2=-A\D T^2+A^{-1}\D r^2+r^2\D l_{n-2}^2,\\
&A(r)=-\frac{2M}{r^{n-3}}-\frac{2\Lambda}{(n-1)(n-2)}r^2,\\
&\rho=-p=\frac{\Lambda}{\kappa_n},
\end{aligned} 
\end{align} 
where the cosmological constant $\Lambda$ has been identified as
\begin{align}
\Lambda\equiv -(n-1)(n-2)h_1 M.
\end{align} 
Hereafter, we assume $\chi\ne -1,0,1$ and consider the domain $r>0$ of the Gamboa solution (\ref{ansatz00-2}).

With $h_1=0$, the Gamboa solution is identical to the topological Schwarzschild-Tangherlini (S-T) vacuum solution
\begin{align}
&\D s^2=-\biggl(-\frac{2M}{r^{n-3}}\biggl)\D T^2+\biggl(-\frac{2M}{r^{n-3}}\biggl)^{-1}\D r^2+r^2\D l_{n-2}^2.\label{Schwarzschild}
\end{align} 
Based on the behavior of this vacuum solution, we refer to a spacetime described by the metric (\ref{ansatz00-2}) as {\it asymptotically topological S-T} if 
\begin{align}
A\simeq { O}(r^{-(n-3)}), \quad B^{-1}\simeq {O}(r^{-(n-3)}) \label{asym-ST}
\end{align} 
as $r\to \infty$.
The conditions (\ref{asym-ST}) imply the asymptotically locally flat condition $\lim_{r\to\infty}{R^{\mu\nu}}_{\rho\sigma}=0$; however, the inverse is not always true.
The solution (\ref{ansatz00-2}) with $h_1\ne 0$ is asymptotically topological S-T only for $\beta\in(2/(n-3),\infty)$, which corresponds to $\chi\in(\chi_0,0)$.
We note that the Gamboa solution for $\chi=\chi_0$ with $A_1\ne 0$ is not asymptotically topological S-T.

In the metric of the Gamboa solution (\ref{ansatz00-2}), the powers of $\Pi$ are integers only when $1/(2+\beta)(=-2\chi/(1-\chi))$ is an integer.
Hence, if $1/(2+\beta)$ is not an integer, the metric is not real in general in the domains where $\Pi<0$ holds.
In such cases, one may use the Gamboa solution in the following form
\begin{align}
\label{ansatz00-2-interior}
\begin{aligned} 
&\D s^2=-A\D T^2+B\D r^2+r^2\D l_{n-2}^2,\\
&A(r)=\frac{2{\bar M}}{r^{n-3}}(-\Pi)^{1/(2+\beta)},\\
&B(r)=\frac{r^{n-3}}{2{\bar M}}(-\Pi)^{-(3+2\beta)/(2+\beta)},\\
&\rho=\frac{p}{\chi}=\frac{\alpha{\bar M}}{r^{(n-3)(1+\beta)}}(-\Pi)^{(1+\beta)/(2+\beta)},
\end{aligned} 
\end{align} 
which is obtained from the solution (\ref{ansatz00-2}) by a coordinate transformation $(-1)^{-(1+\beta)/(2+\beta)}T\to T$ and a reparametrization $-(-1)^{(3+2\beta)/(2+\beta)}M\to {\bar M}$.

\subsection{Energy conditions for the matter field}

The Gamboa spacetime (\ref{ansatz00-2}) for $M<(>)0$ is static (dynamical), while the spacetime given by Eq.~(\ref{ansatz00-2-interior}) for ${\bar M}<(>)0$ is dynamical (static).
The coordinates $T$ and $r$ are spacelike and timelike in the dynamical region, respectively.
As a consequence, the matter field in the dynamical region is not a perfect fluid but rather an anisotropic fluid with the energy density $\mu$, radial pressure $p_1$, and tangential pressure $p_2$ given by
\begin{align}
\label{matter-interior}
&\mu=-p, \qquad p_1=-\rho,\qquad p_2=p
\end{align} 
with $p$ and $\rho$ given by Eq.~(\ref{ansatz00-2}) or (\ref{ansatz00-2-interior})~\cite{Maeda:2024tpl}.
Alternatively, the matter field in the dynamical region may be interpreted as a {\it spacelike} (or {\it tachyonic}) perfect fluid~\cite{Maeda:2024tpl}.

The standard energy conditions consist of the {\it null} energy condition (NEC), {\it weak} energy condition (WEC), {\it dominant} energy condition (DEC), and {\it strong} energy condition (SEC)~\cite{Maeda:2018hqu}.
Equivalent representations of these conditions for a perfect fluid obeying $p=\chi\rho$ are given by
\begin{align}
\mbox{NEC}:&~~(1+\chi)\rho\ge 0,\label{NEC-I}\\
\mbox{WEC}:&~~\rho\ge 0\mbox{~in addition to NEC},\label{WEC-I}\\
\mbox{DEC}:&~~(1-\chi)\rho\ge 0\mbox{~in addition to WEC},\label{DEC-I}\\
\mbox{SEC}:&~~[(n-3)+(n-1)\chi]\rho\ge 0 \nonumber \\
&~~\mbox{~in addition to NEC}.\label{SEC-I}
\end{align}
For the matter field in the dynamical region, by Eq.~(\ref{matter-interior}), equivalent representations are given by 
\begin{align}
\mbox{NEC}:&~~(1+\chi)\rho\le 0,\label{NEC-I-interior}\\
\mbox{WEC}:&~~\chi \rho\le 0\mbox{~in addition to NEC},\label{WEC-I-interior}\\
\mbox{DEC}:&~~(1-\chi)\rho\ge 0\mbox{~in addition to WEC},\label{DEC-I-interior}\\
\mbox{SEC}:&~~(1-\chi)\rho\le 0\mbox{~in addition to NEC}.\label{SEC-I-interior}
\end{align}

\subsection{Killing horizon}
\label{sec:killing}

With $h_1>0$ for $\chi\ne 0,\chi_0,1$ and $h_1\ne 0$ for $\chi=\chi_0$, $\Pi=0$ admits a real solution $r=r_{\rm h}(>0)$ given by 
\begin{align}
\label{def-rh}
&r_{\rm h}:=\left\{
\begin{array}{ll}
h_1^{1/[(n-3)\beta-2]} & [\chi\ne \chi_0,1],\\
e^{1/h_1} & [\chi=\chi_0],
\end{array}
\right.
\end{align} 
which satisfies $A(r_{\rm h})=0=\Pi(r_{\rm h})$.
The coordinate systems (\ref{ansatz00-2}) and (\ref{ansatz00-2-interior}) are singular at $r=r_{\rm h}$ since $B$ diverges there.

To remove this coordinate singularity $r=r_{\rm h}$ in the coordinate system (\ref{ansatz00-2}), we first introduce the following quasiglobal coordinates
\begin{align}
\D s^2=&-H(x)\D t^2+H(x)^{-1}\D x^2+r(x)^2\D l_{n-2}^2, \label{metric-Buchdahl}
\end{align}
using the following coordinate transformation and identification:
\begin{align}
&T=\varepsilon \Omega t,\label{trans-t}\\
&x=\Omega\int \Pi(r)^{-(1+\beta)/(2+\beta)}\D r,\label{trans1}\\
&H(x)\equiv \Omega^2 A(r(x)),\label{trans2}
\end{align} 
where a constant $\Omega$ has been introduced as a gauge choice and $\varepsilon =\pm 1$ is chosen such that $\D T/\D t>1$.
For the solution (\ref{ansatz00-2-interior}), we define the coordinate $x$ by 
\begin{align}
&x=\Omega\int (-\Pi)^{-(1+\beta)/(2+\beta)}\D r \label{trans1-interior}
\end{align} 
instead of Eq.~(\ref{trans1}).
We have $\D r/\D x>(<)0$ for $\Omega>(<)0$.
In the quasiglobal coordinates, $r=r_{\rm h}$ corresponds to $x=x_{\rm h}$ determined by $H(x_{\rm h})=0$, which is still a coordinate singularity.

The coordinate singularity at $x=x_{\rm h}$ can finally be removed by introducing an ingoing-null coordinate $v:=t+\int H(x)^{-1}\D x$, with which the metric (\ref{metric-Buchdahl}) is written as
\begin{equation}
\D s^2=-H(x)\D v^2+2\D v\D x+r(x)^2\D l_{n-2}^2.\label{metric-Buchdahl-v}
\end{equation}
One may use an outgoing-null coordinate $u:=t-\int H(x)^{-1}\D x$, and the resulting metric is given by Eq.~(\ref{metric-Buchdahl-v}) with $v$ replaced by $-u$.
If $x=x_{\rm h}$ is regular and $r(x_{\rm h})\ne 0$ holds in the single-null coordinates (\ref{metric-Buchdahl-v}), it is a Killing horizon associated with a Killing vector $\xi^\mu(\partial/\partial x^\mu)=\partial/\partial v$ which is a null hypersurface.

In general relativity, a $C^{1,1}$ metric, often denoted by $C^{2-}$ in physics, is sufficient for regularity, as it avoids curvature singularities as well as the divergence of matter fields.
(See Sec.~2.3 in Ref.~\cite{Maeda:2024lbq}.)
We note that the $C^{1,1}$ regularity avoids a massive thin shell which is a curvature singularity described by the $C^{0,1}$ metric.
Actually, by Proposition~2 in Ref.~\cite{Maeda:2024lbq}, $r=r_{\rm h}$, corresponding to $x=x_{\rm h}$, in the Gamboa solution given by Eq.~(\ref{ansatz00-2}) or (\ref{ansatz00-2-interior}) is not regular but instead a {\it parallelly propagated curvature singularity} unless $-1/3\le \chi<0$ or $\chi=-1$.

In fact, according to Proposition~6 in Ref.~\cite{Maeda:2024lbq}, $r=r_{\rm h}$ is a nondegenerate Killing horizon in the regular coordinate system (\ref{metric-Buchdahl-v}) for $\chi\in[-1/3,0)$ or, equivalently, $\beta\in[0,\infty)$.
For this reason, we will assume $\chi\in[-1/3,0)$ in the following sections.
As we will see in Sec.~\ref{sec:infinity}, the Killing horizon $r=r_{\rm h}$ corresponds to extendible null boundaries in the Penrose diagram.
According to Proposition~8 in Ref.~\cite{Maeda:2024lbq}, a matter field exists on the horizon only for $\chi=-1/3$.
(See also Proposition~2 in Ref.~\cite{Maeda:2021ukk}.)
We present an explicit form of the matter field confined on the Killing horizon in the Gamboa solution for $\chi=-1/3$ in Appendix~\ref{app:geodesic}.

\subsection{Asymptotic region $r\to \infty$ and singularity $r=0$ for $\chi\in[-1/3,0)$}
\label{sec:infinity}

In the Gamboa solution, $r=0$ corresponds to a curvature singularity since $\rho$ blows up there.
Now we study the boundaries $r\to \infty$ and $r=0$ of the Gamboa solution in the forms of Eqs.~(\ref{ansatz00-2}) and (\ref{ansatz00-2-interior}) for $\chi\in[-1/3,0)$.

The causal nature of boundaries is determined by the two-dimensional Lorentzian portion of the line element (\ref{ansatz00-2}), which is written in the conformally flat form as
\begin{align}
&\D s^2=A(r)(-\D T^2+\D r_*^2),\\
&r_*(r):=\int^r\sqrt{\frac{B}{A}}\D r.\label{def-r*}
\end{align}
If $r_*$ converges to a finite value as $r\to r_0$ in a region where $g_{TT}<(>)0$ holds, $r=r_0$ corresponds to timelike (spacelike) hypersurface in the Penrose diagram, while it corresponds to a null boundary in the diagram if $\lim_{r\to r_0}r_*$ blows up.
Near $r\to \infty$, we obtain
\begin{align}
&\lim_{r\to \infty}\sqrt{\frac{B}{A}} \nonumber\\
&\propto \left\{
\begin{array}{ll}
r^{(n-3)(1+\beta)-2} & [\chi\in[-1/3,\chi_0)],\\
r^{n-3}/\ln|r|& [\chi=\chi_0],\\
r^{n-3} & [\chi=(\chi_0,0)],
\end{array}
\right.
\end{align}
which shows $\lim_{r\to \infty}r_*\to \infty$ for $\chi\in(\chi_0,0)$.
This is also true for $\chi=\chi_0$ shown by 
\begin{align}
&\lim_{r\to \infty}\int^r\frac{r^{n-3}}{\ln|r|}\D r >\lim_{r\to \infty}\int^rr^{n-4}\D r\to \infty.
\end{align}
Also, $\lim_{r\to \infty}r_*\to \infty$ is satisfied for $\chi\in[-1/3,\chi_0)$ because the inequality $(n-3)(1+\beta)-2\ge -1$ is equivalent to
\begin{align}
\chi\ge -\frac{n-3}{n-1},\label{key-ineq}
\end{align}
which is satisfied for $\chi\in[-1/3,\chi_0)$.
To summarize, $r\to \infty$ is causally null in the Penrose diagram for $\chi\in[-1/3,0)$.

On the other hand, near $r\to 0$, we obtain
\begin{align}
&\lim_{r\to 0}\sqrt{\frac{B}{A}} \nonumber\\
&\propto \left\{
\begin{array}{ll}
r^{n-3} & [\chi\in[-1/3,\chi_0)],\\
r^{n-3}/\ln|r| & [\chi=\chi_0],\\
r^{(n-3)(1+\beta)-2} & [\chi=(\chi_0,0)],
\end{array}
\right.
\end{align}
which shows that $r_*$ converges to a finite value as $r\to 0$ for $\chi\in[-1/3,\chi_0)$ and $\chi=(\chi_0,0)$, where $(n-3)(1+\beta)-2\ge n-3$ holds in the latter case.
This is also true for $\chi=\chi_0$ shown by
\begin{align}
&\lim_{r\to 0}\int^r\frac{r^{n-3}}{\ln|r|}\D r <\lim_{r\to 0}\int^rr^{n-4}\D r<\infty.
\end{align}
Thus, $r=0$ is causally non-null in the Penrose diagram for $\chi\in[-1/3,0)$.

To identify null infinities, we study an affinely parametrized radial null geodesic $\gamma$ in the spacetime given by Eq.~(\ref{ansatz00-2}) with its tangent vector $k^\mu=(\D T/\D\lambda,\D r/\D\lambda,0,\cdots,0)$, where $\lambda$ is an affine parameter along $\gamma$.
Using a conserved quantity $E:=-\xi_\mu k^\mu=Ak^{T}$ along $\gamma$, where $\xi^\mu=(1,0,\cdots,0)$ is a hypersurface-orthogonal Killing vector, and the null condition $k_\mu k^\mu=0$, we obtain
\begin{align}
\lambda(r)=\pm E^{-1}\int^r\sqrt{AB}\D r \label{null-infinity}
\end{align}
for $E\ne 0$.
Here the sign implies two possible directions of $\gamma$. 
While $r=r_0$ is null infinity if $\lim_{r\to r_0}\lambda$ blows up, $r=r_0$ is extendible if it is regular and $\lim_{r\to r_0}\lambda$ is finite.
Near $r\to \infty$, we obtain
\begin{align}
&\lim_{r\to \infty}\sqrt{AB} \nonumber\\
&\propto \left\{
\begin{array}{ll}
r^{-[2-(n-3)\beta](1+\beta)/(2+\beta)} & [\chi\in[-1/3,\chi_0)],\\
(\ln|r|)^{-(n-1)/[2(n-2)]} & [\chi=\chi_0],\\
1 & [\chi\in(\chi_0,0)],
\end{array}
\right.
\end{align}
which shows that $\lim_{r\to \infty}\lambda\to \infty$ for $\chi\in(\chi_0,0)$.
This is also true for $\chi=\chi_0$ shown by
\begin{align}
&\lim_{r\to \infty}\int^r(\ln|r|)^{-(n-1)/[2(n-2)]}\D r \nonumber\\
&>\lim_{r\to \infty}\int^rr^{-(n-1)/[2(n-2)]}\D r\to \infty.
\end{align}
Also, $\lim_{r\to \infty}\lambda\to \infty$ is satisfied for $\chi\in[-1/3,\chi_0)$ because the inequality $-[2-(n-3)\beta](1+\beta)/(2+\beta)\ge -1$ is equivalent to Eq.~(\ref{key-ineq}), which is satisfied for $\chi\in[-1/3,\chi_0)$.
To summarize, $r\to \infty$ is null infinity for $\chi\in[-1/3,0)$.

On the other hand, near $r\to 0$, we obtain
\begin{align}
&\lim_{r\to 0}\sqrt{AB} \nonumber\\
&\propto \left\{
\begin{array}{ll}
1 & [\chi\in[-1/3,\chi_0)],\\
(\ln|r|)^{-(n-1)/[2(n-2)]} & [\chi=\chi_0],\\
r^{[(n-3)\beta-2](1+\beta)/(2+\beta)} & [\chi\in(\chi_0,0)],
\end{array}
\right.
\end{align}
which shows that $r_*$ converges to a finite value as $r\to 0$ for $\chi\in[-1/3,\chi_0)$ and $\chi=(\chi_0,0)$, where $(n-3)(1+\beta)-2\ge n-3$ holds in the latter case.
This is also true for $\chi=\chi_0$ shown by
\begin{align}
&\lim_{r\to 0}\int^r(\ln|r|)^{-(n-1)/[2(n-2)]}\D r \nonumber\\
&<\lim_{r\to 0}\int^rr^{-(n-1)/[2(n-2)]}\D r< \infty.
\end{align}
Thus, $r=0$ is not null infinity for $\chi\in[-1/3,0)$.

\section{Extension beyond the Killing horizon}
\label{sec:extension}

In this section, we assume $\chi\in[-1/3,0)$ or, equivalently, $\beta\in[0,\infty)$ and consider the parameter space where $\Pi=0$ admits a solution $r=r_{\rm h}$ given by Eq.~(\ref{def-rh}).
We will explicitly show that $r=r_{\rm h}$ is a nondegenerate Killing horizon and the spacetime can be extended beyond there in the at least $C^{1,1}$ regular manner.
In general, closed-form expressions of the integrals~(\ref{trans1}) and (\ref{trans1-interior}) are not available.
However, they are available with $\beta=4/(n-4)$ or, equivalently, $\chi=-(n-4)/(3n-4)$ for $n\ge 5$.
We first discuss this special case and then deal with the most general case.

\subsection{For $\chi=-(n-4)/(3n-4)$ with $n\ge 5$}
\label{sec:example-exact}

In the special case of $\chi=-(n-4)/(3n-4)(>\chi_0)$ with $n\ge 5$, the Gamboa solution given by Eq.~(\ref{ansatz00-2}) or (\ref{ansatz00-2-interior}) can be written in the single-null coordinates (\ref{metric-Buchdahl-v}) with the metric represented by elementary functions.
The resulting new form of the Gamboa solution is given by 
\begin{align}
\label{Gamboa-H}
\begin{aligned}
&r(x)=\left(h_1+m\Delta^{2+\beta}\right)^{1/(2+\beta)},\\
&H(x)=\frac{2\Delta}{r^{2(2+\beta)/\beta}}\biggl(=\frac{2\Delta}{r^{n-2}}\biggl),\\
&\rho(x)=\frac{p}{\chi}=-\frac{\alpha m\Delta^{1+\beta}}{r^{2(1+\beta)(2+\beta)/\beta}},
\end{aligned} 
\end{align} 
where $\Delta:=x-x_{\rm h}$ and $x_{\rm h}$ is an integration constant of the integral~(\ref{trans1}) or (\ref{trans1-interior}) such that $r=r_{\rm h}(=h_1^{1/(2+\beta)})$ corresponds to $x=x_{\rm h}$.
In fact, although we will keep $x_{\rm h}$ nonvanishing in the following discussions, it is not a physical parameter as we can set $x_{\rm h}=0$ by a shift transformation $x\to x+x_{\rm h}$.
Therefore, this new form is parametrized by two real constants $h_1$ and $m$, and we assume that $h_1> 0$.

While the new form (\ref{Gamboa-H}) is valid across the horizon $x=x_{\rm h}$ for an integer $\beta$, it is valid only in the domain $x\ge x_{\rm h}$ for a noninteger $\beta$.
For a noninteger $\beta$, the domain $x\le x_{\rm h}$ is described by the single-null coordinates (\ref{metric-Buchdahl-v}) with 
\begin{align}
\label{Gamboa-H-interior}
\begin{aligned}
&r(x)=\left[h_1-{\bar m}(-\Delta)^{2+\beta}\right]^{1/(2+\beta)},\\
&H(x)=\frac{2\Delta}{r^{2(2+\beta)/\beta}}\biggl(=\frac{2\Delta}{r^{n-2}}\biggl),\\
&\rho(x)=\frac{p}{\chi}=-\frac{\alpha{\bar m}(-\Delta)^{1+\beta}}{r^{2(1+\beta)(2+\beta)/\beta}}.
\end{aligned} 
\end{align} 
This form of the solution is parametrized by two real constants $h_1$ and ${\bar m}$.
Since the solution (\ref{Gamboa-H-interior}) with an even (odd) $\beta$, which is realized only for $n=5,6$ ($n=8$), is identical to the solution (\ref{Gamboa-H}) with $m=-{\bar m}$ ($m={\bar m}$), we assume that $\beta$ is a noninteger and then the domain of the solution given by Eq.~(\ref{Gamboa-H-interior}) is $x\le x_{\rm h}$.

We note that the solution~(\ref{Gamboa-H}) with $h_1>0$ and $m=0$ and the solution (\ref{Gamboa-H-interior}) with $h_1>0$ and ${\bar m}=0$ are locally Minkowski.
It is emphasized that the Minkowski spacetime is not realized in the parameter space of the Gamboa solution in the original forms given by Eqs.~(\ref{ansatz00-2}) and (\ref{ansatz00-2-interior}).
Hereafter, we assume that $m\ne 0$ and ${\bar m}\ne 0$ and clarify the relations between the new forms given by Eqs.~(\ref{Gamboa-H}) and (\ref{Gamboa-H-interior}) to the original forms given by Eqs.~(\ref{ansatz00-2}) and (\ref{ansatz00-2-interior}).
\begin{table*}[htb]
\begin{center}
\caption{The correspondence between the Gamboa solution in the forms of Eqs.~(\ref{Gamboa-H}) and (\ref{Gamboa-H-interior}) in the quasiglobal coordinates (\ref{metric-Buchdahl}) for $\chi=-(n-4)/(3n-4)$ with $n\ge 5$ and the forms of Eqs.~(\ref{ansatz00-2}) and (\ref{ansatz00-2-interior}) with $h_1>0$.}
\label{table:correspondence1}
\begin{tabular}{|c|c|c|c|c|}\hline
Metric & $\beta$ & $m,{\bar m}$ & $x<x_{\rm h}$ & $x>x_{\rm h}$ \\ \hline
Eq.~(\ref{Gamboa-H}) & Even& $m>0$ & Eq.~(\ref{ansatz00-2}) with $M>0 $ & Eq.~(\ref{ansatz00-2}) with $M< 0$ \\ \cline{3-5}
& ($n=5,6$) & $m<0$ & Eq.~(\ref{ansatz00-2-interior}) with ${\bar M}<0$ & Eq.~(\ref{ansatz00-2-interior}) with ${\bar M}> 0$ \\ \cline{2-5}
& Odd & $m>0$ & Eq.~(\ref{ansatz00-2-interior}) with ${\bar M}<0$ & Eq.~(\ref{ansatz00-2}) with $M<0$ \\ \cline{3-5}
& ($n=8$) & $m<0$ & Eq.~(\ref{ansatz00-2}) with $M>0$ & Eq.~(\ref{ansatz00-2-interior}) with ${\bar M}>0$ \\ \cline{2-5}
& Noninteger & $m>0$ & not applicable & Eq.~(\ref{ansatz00-2}) with $M<0$ \\ \cline{3-5}
& ($n\ne 5,6,8$) & $m<0$ & not applicable & Eq.~(\ref{ansatz00-2-interior}) with ${\bar M}>0$ \\ \hline
Eq.~(\ref{Gamboa-H-interior}) & Noninteger & ${\bar m}>0$ & Eq.~(\ref{ansatz00-2-interior}) with ${\bar M}<0$ & not applicable \\ \cline{3-5}
& ($n\ne 5,6,8$) & ${\bar m}<0$ & Eq.~(\ref{ansatz00-2}) with $M>0$ & not applicable \\ \hline
\end{tabular}
\end{center}
\end{table*}

By coordinate transformations
\begin{align}
\label{x-r-exact}
\begin{aligned}
&t=- \varepsilon MT,\\
&x=x_{\rm h}- M^{-1}(r^{2+\beta}-h_1)^{1/(2+\beta)} 
\end{aligned} 
\end{align} 
with $m\equiv (- M)^{2+\beta}$ from the quasiglobal coordinates (\ref{metric-Buchdahl}) with Eq.~(\ref{Gamboa-H}), we obtain the Gamboa solution in the form of Eq.~(\ref{ansatz00-2}).
For an integer $\beta$, the domain $x> (< )x_{\rm h}$ is described by the solution (\ref{ansatz00-2}) with $M<(>)0$ by Eq.~(\ref{x-r-exact}).
In particular, $m>0$ holds for $M\ne 0$ with an even $\beta$, while $m>(<)0$ holds for $M<(>)0$ with an odd $\beta$.
For a noninteger $\beta$, the domain $x> x_{\rm h}$ with $m>0$ is described by the solution (\ref{ansatz00-2}) with $M<0$.

On the other hand, by coordinate transformations
\begin{align}
\label{x-r-exact-interior}
\begin{aligned}
&t=-\varepsilon {\bar M}T,\\
&x= x_{\rm h}+{\bar M}^{-1}(h_1-r^{2+\beta})^{1/(2+\beta)}
\end{aligned} 
\end{align} 
with $m\equiv -{\bar M}^{2+\beta}$, we obtain the Gamboa solution in the form of Eq.~(\ref{ansatz00-2-interior}).
For an integer $\beta$, the domain $x> (<)x_{\rm h}$ is described by the solution (\ref{ansatz00-2-interior}) with ${\bar M}>(<)0$ by Eq.~(\ref{x-r-exact-interior}).
In particular, $m<0$ holds for ${\bar M}\ne 0$ with an even $\beta$, while $m>(<)0$ holds for ${\bar M}<(>)0$ with an odd $\beta$.
For a noninteger $\beta$, the domain $x\ge x_{\rm h}$ with $m<0$ is described by the solution (\ref{ansatz00-2-interior}) with ${\bar M}>0$.

By the coordinate transformations (\ref{x-r-exact}) with ${\bar m}:=-M^{2+\beta}$ from the quasiglobal coordinates (\ref{metric-Buchdahl}) with Eq.~(\ref{Gamboa-H-interior}), we obtain the Gamboa solution in the form of Eq.~(\ref{ansatz00-2}).
For a noninteger $\beta$, the domain $x<x_{\rm h}$ with ${\bar m}<0$ is described by the solution (\ref{ansatz00-2}) with $M>0$.
On the other hand, using the coordinate transformations (\ref{x-r-exact-interior}) with ${\bar m}:=(-{\bar M})^{2+\beta}$, we obtain the Gamboa solution in the form of Eq.~(\ref{ansatz00-2-interior}).
For a noninteger $\beta$, the domain $x<x_{\rm h}$ with ${\bar m}>0$ is described by the solution~(\ref{ansatz00-2-interior}) with ${\bar M}<0$.
The results are summarized in Table~\ref{table:correspondence1}.

Suppose that the parameters in the domains $x\ge x_{\rm h}$ and $x\le x_{\rm h}$ are given by $(h_1,m)=(h_1^+,m_+)$ and $(h_1,m)=(h_1^-,m_-)$, respectively.
Then, only a single condition $h_1^+=h_1^-$ guarantees the $C^{1,1}$ regularity at the horizon $x=x_{\rm h}$ because the values of $\chi$ and $\beta$ are fixed with a given number of $n(\ge 5)$ in the special case under consideration.
In particular, we can set $m_+=0$ ($m_-=0$) with $m_-\ne 0$ ($m_+\ne 0$), and the domain $x<(>)x_{\rm h}$ is then flat.
With $m_+=0$ and $m_-\ne 0$, the maximally extended spacetime describes a curious black hole with the Minkowski exterior~\cite{Maeda:2025ghc}.

\subsection{For general $\chi\in[-1/3,0)$}

Although the right-hand side of Eq.~(\ref{trans1}) can be written by the hypergeometric function, the explicit form of $r=r(x)$ cannot be obtained in the general case.
Nevertheless, as given in the proof of Proposition~6 in Ref.~\cite{Maeda:2024lbq}, the asymptotic solution near the Killing horizon $x=x_{\rm h}$ in the domain $x> x_{\rm h}$ is given by
\begin{align}
\label{Asymp-sol}
\begin{aligned}
&H(x)\simeq H_1\Delta+H_{3+\beta}\Delta^{3+\beta},\\
&r(x)\simeq r_{\rm h}+r_{2+\beta}\Delta^{2+\beta},\\
&\rho\simeq -\frac{(n-2)(2+\beta)(1+\beta)H_1r_{2+\beta}}{\kappa_n(1+\chi)r_{\rm h}}\Delta^{1+\beta},
\end{aligned}
\end{align}
where $\Delta:=x-x_{\rm h}$, $r_{\rm h}= {h_1}^{1/[(n-3)\beta-2]}$ and $H_1\ne 0$.
The coefficient $H_{3+\beta}$ is given by 
\begin{align}
H_{3+\beta}=-\frac{H_1r_{2+\beta}[2(n-3)\beta+ 3n-10]}{(3+\beta)r_{\rm h}}.\label{H3+beta-out-main}
\end{align}
In fact, Eq.~(\ref{Asymp-sol}) with $r_{2+\beta}=H_{3+\beta}=0$ and $H_1\ne 0$ is not an asymptotic solution but the {\it exact} Minkowski solution defined in the domain $-\infty<x<\infty$, where $x>x_{\rm h}$ and $x<x_{\rm h}$ correspond to the Rindler (Milne) and Milne (Rindler) charts for $H_1>(<)0$, respectively, and $x=x_{\rm h}$ is the Rindler horizon~\cite{Maeda:2025ghc}.
We note that the asymptotic solution (\ref{Asymp-sol}) with $r_{2+\beta}\ne 0$ is not valid in the domain $x<x_{\rm h}$ unless $\beta$ is an integer.

First, we clarify the domains where the Gamboa solution in the forms of Eqs.~(\ref{ansatz00-2}) and (\ref{ansatz00-2-interior}) are mapped in the quasiglobal coordinates (\ref{metric-Buchdahl}) (and therefore in the single-null coordinates (\ref{metric-Buchdahl-v})).
\begin{lm}
\label{lm:Gamboa-quasi-global}
Suppose that the Gamboa solution for $\chi\in[-1/3,0)$ or, equivalently, $\beta\in[0,\infty)$ admits a real solution $r=r_{\rm h}$ to $\Pi(r)=0$ given by Eq.~(\ref{def-rh}).
Then the solution in the forms of Eq.~(\ref{ansatz00-2}) with $M<(>)0$ and Eq.~(\ref{ansatz00-2-interior}) with ${\bar M}>(<)0$ are mapped onto the domain $x>(<)x_{\rm h}$ in the quasiglobal coordinates (\ref{metric-Buchdahl}).
\end{lm}
{\it Proof}. 
Near $r=r_{\rm h}$, Eq.~(\ref{def-Pi}) gives
\begin{align}
\label{def-Pi-near}
&\Pi(r)\simeq \Pi_1(r-r_{\rm h}),
\end{align} 
where
\begin{align}
\label{def-Pi1}
&\Pi_1:=\left\{
\begin{array}{ll}
[(n-3)\beta-2]/r_{\rm h} & [\chi\ne \chi_0],\\
-h_1/r_{\rm h} & [\chi=\chi_0].
\end{array}
\right.
\end{align} 
Hence, $\Pi>(<)0$ holds in the domains $r>(<)r_{\rm h}$ for $\Pi_1>0$ and $r<(>)r_{\rm h}$ for $\Pi_1<0$.

For the Gamboa solution in the form of Eq.~(\ref{ansatz00-2}) defined in the domain where $\Pi>0$ holds, Eq.~(\ref{trans1}) gives
\begin{align}
&x-x_{\rm h}\simeq (2+\beta)\Omega\Pi_1^{-1}[\Pi_1(r-r_{\rm h})]^{1/(2+\beta)}\label{x-r1}
\end{align} 
and therefore the solution is mapped onto the domain $x>(<)x_{\rm h}$ for $\Omega/\Pi_1>(<)0$.
Equations~(\ref{trans-t})--(\ref{trans2}) with Eq.~(\ref{x-r1}) give
\begin{align}
\label{asymp-H1-generalOmega}
\begin{aligned}
&r\simeq r_{\rm h}+\frac{1}{\Pi_1(2+\beta)^{2+\beta}}(\Pi_1\Omega^{-1}\Delta)^{2+\beta},\\
&H\simeq-\frac{2\Omega \Pi_1M}{(2+\beta)r_{\rm h}^{n-3}}\Delta+O(\Delta^{3+\beta}),\\
&\rho\simeq \frac{\alpha M}{(2+\beta)^{1+\beta}r_{\rm h}^{(n-3)(1+\beta)}} (\Pi_1\Omega^{-1}\Delta)^{1+\beta}.
\end{aligned} 
\end{align} 
Setting $\Omega=-(2+\beta)/(M\Pi_1)$ as a gauge choice, we obtain
\begin{align}
\label{asymp-H1}
\begin{aligned}
&r\simeq r_{\rm h}+\frac{(-M\Pi_1^2\Delta)^{2+\beta}}{\Pi_1(2+\beta)^{2(2+\beta)}},\\
&H\simeq\frac{2}{r_{\rm h}^{n-3}}\Delta+O(\Delta^{3+\beta}),\\
&\rho\simeq \frac{\alpha M(-M\Pi_1^2\Delta)^{1+\beta}}{(2+\beta)^{2(1+\beta)}r_{\rm h}^{(n-3)(1+\beta)}}.
\end{aligned} 
\end{align} 
Thus, using $\Omega/\Pi_1=-(2+\beta)/(M\Pi_1^2)$, the Gamboa solution in the form of Eq.~(\ref{ansatz00-2}) with $M<(>)0$ is mapped onto the domain $x>(<)x_{\rm h}$, where the asymptotic expansion (\ref{asymp-H1}) is valid even with a noninteger $\beta$.

For the Gamboa solution in the form of Eq.~(\ref{ansatz00-2-interior}) defined in the domain where $\Pi<0$ holds, Eq.~(\ref{trans1-interior}) gives
\begin{align}
&x-x_{\rm h}\simeq -(2+\beta)\Omega\Pi_1^{-1} [-\Pi_1(r-r_{\rm h})]^{1/(2+\beta)},\label{x-r2}
\end{align} 
and therefore the solution is mapped onto the domain $x>(<)x_{\rm h}$ for $\Omega/\Pi_1<(>)0$.
Equations~(\ref{trans-t}), (\ref{trans2}), and~(\ref{x-r2}) give
\begin{align}
\begin{aligned} 
&r\simeq r_{\rm h}-\frac{1}{\Pi_1(2+\beta)^{2+\beta}}(-\Pi_1\Omega^{-1}\Delta)^{2+\beta},\\
&H\simeq -\frac{2\Omega\Pi_1{\bar M}}{(2+\beta)r_{\rm h}^{n-3}}\Delta+O(\Delta^{3+\beta}),\\
&\rho\simeq \frac{\alpha{\bar M}}{(2+\beta)^{1+\beta}r_{\rm h}^{(n-3)(1+\beta)}}(-\Pi_1\Omega^{-1}\Delta)^{1+\beta}.
\end{aligned} 
\end{align} 
Setting $\Omega=-(2+\beta)/({\bar M}\Pi_1)$ as a gauge choice, we obtain
\begin{align}
\label{asymp-H1-1}
\begin{aligned}
&r\simeq r_{\rm h}-\frac{({\bar M}\Pi_1^2\Delta)^{2+\beta}}{\Pi_1(2+\beta)^{2(2+\beta)}},\\
&H\simeq \frac{2}{r_{\rm h}^{n-3}}\Delta+O(\Delta^{3+\beta}),\\
&\rho\simeq \frac{\alpha{\bar M}({\bar M}\Pi_1^2\Delta)^{1+\beta}}{(2+\beta)^{2(1+\beta)}r_{\rm h}^{(n-3)(1+\beta)}}.
\end{aligned} 
\end{align} 
Thus, using $\Omega/\Pi_1=-(2+\beta)/({\bar M}\Pi_1^2)$, the Gamboa solution in the form of Eq.~(\ref{ansatz00-2-interior}) with ${\bar M}>(<)0$ is mapped into the domain $x>(<)x_{\rm h}$, where the asymptotic expansion (\ref{asymp-H1-1}) is valid even with a noninteger $\beta$.
\qed
\bigskip

The following proposition shows how to extend the Gamboa solution in the form of Eq.~(\ref{ansatz00-2}) or (\ref{ansatz00-2-interior}) beyond the Killing horizon $r=r_{\rm h}$ in the single-null coordinates (\ref{metric-Buchdahl-v}) for $\chi\in[-1/3,0)$.
We emphasize that the extended region can be described by the Gamboa solution even for a different value of $\chi$.
(See Ref.~\cite{addendum} for a refined statement of Proposition~6 in Ref.~\cite{Maeda:2024lbq}.)
\begin{Prop}
\label{Pro:Gamboa-extension1}
Let $({\cal M}_n^+,g_{\mu\nu}^+)$ and $({\cal M}_n^-,g_{\mu\nu}^-)$ be spacetimes described by the Gamboa solution in the form of Eq.~(\ref{ansatz00-2}) or (\ref{ansatz00-2-interior}) for $\chi=\chi_+$ ($\beta=\beta_+$) and $\chi= \chi_-$ ($\beta=\beta_-$) defined in the domains $x>x_{\rm h}$ and $x<x_{\rm h}$ in the quasiglobal coordinates (\ref{metric-Buchdahl}), respectively, according to Lemma~\ref{lm:Gamboa-quasi-global}.
If $r(x)$ is continuous at $x=x_{\rm h}$, $x=x_{\rm h}$ is a nondegenerate Killing horizon.
Then the metric in the single-null coordinates (\ref{metric-Buchdahl-v}) is $C^\infty$ at the horizon only for $\chi_+=\chi_-=-1/(1+2N)$ with $N\in{\mathbb N}$ or, equivalently, $\beta_+=\beta_- \in{\mathbb N}_0$, where the domains $x>x_{\rm h}$ and $x<x_{\rm h}$ are described by the Gamboa solutions as shown in Table~\ref{table:C-infty}.
\end{Prop}
{\it Proof}. 
By Eq.~(\ref{def-rh}), the continuity of $r(x)$ at the horizon requires
\begin{align}
&(h_1^+)^{1/[(n-3)\beta_+-2]}=(h_1^-)^{1/[(n-3)\beta_--2]} \label{continuous-r1}
\end{align} 
in the case of $\chi_+\ne \chi_0$ and $\chi_-\ne \chi_0$,
\begin{align}
&(h_1^\pm)^{1/[(n-3)\beta_\pm-2]}=e^{1/h_1^\mp} \label{continuous-r2}
\end{align} 
in the case of $\chi_\pm\ne \chi_0$ with $\chi_\mp=\chi_0$, and 
\begin{align}
&h_1^+=h_1^- \label{continuous-r3}
\end{align} 
in the case of $\chi_+=\chi_-=\chi_0$.
The conditions (\ref{continuous-r1})--(\ref{continuous-r3}) can be satisfied by choosing $h_1^+$ and $h_1^-$ appropriately independently of $M$ and ${\bar M}$.
Then the asymptotic expansions~(\ref{asymp-H1}) and (\ref{asymp-H1-1}) show that $r$, $r'$, $H$, and $H'$ are continuous and $r''$ and $H''$ are finite at $x=x_{\rm h}$.
Then, since the metric in the single-null coordinates (\ref{metric-Buchdahl-v}) is at least $C^{1,1}$ at $x=x_{\rm h}$ and $H'\ne 0$ holds, $x=x_{\rm h}$ is a nondegenerate Killing horizon.

The asymptotic expansions~(\ref{asymp-H1}) and (\ref{asymp-H1-1}) also show that the metric at $x=x_{\rm h}$ cannot be $C^\infty$ unless $\beta_+= \beta_-=\beta$ with an integer $\beta$.
For an integer $\beta$, $r(x)$ in Eqs.~(\ref{asymp-H1}) and (\ref{asymp-H1-1}) is written as
\begin{align}
\label{asymp-r}
&r\simeq r_{\rm h}+r_{2+\beta}\Delta^{2+\beta}
\end{align} 
with
\begin{align}
r_{2+\beta}=\frac{(-M)^{2+\beta}\Pi_1^{3+2\beta}}{(2+\beta)^{2(2+\beta)}}
\end{align} 
and
\begin{align}
&r_{2+\beta}=-\frac{{\bar M}^{2+\beta}\Pi_1^{3+2\beta}}{(2+\beta)^{2(2+\beta)}},
\end{align} 
respectively.
As $r_{\rm h}$ and $\beta$ are the same on both sides of the horizon, $\Pi_1$ is the same as well.
Hence, for an integer $\beta$, the metric is at least $C^{2+\beta}$ at the horizon if the domains $x>x_{\rm h}$ and $x<x_{\rm h}$ are described by the Gamboa solutions as shown in Table~\ref{table:C-infty}.
Then, by Proposition~9 in Ref.~\cite{Maeda:2024lbq}, the metric is in fact $C^\infty$ at $x=x_{\rm h}$.
\qed

\begin{table*}[htb]
\begin{center}
\caption{Possible $C^\infty$ attachments at $x=x_{\rm h}$ in the single-null coordinates (\ref{metric-Buchdahl-v}) of the Gamboa solution in the forms of Eq.~(\ref{ansatz00-2}) and (\ref{ansatz00-2-interior}) shown in Proposition~\ref{Pro:Gamboa-extension1}.}
\label{table:C-infty}
\scalebox{1.0}{
\begin{tabular}{|c|c|c|c|c|}\hline
$\beta(=\beta_+=\beta_-)$ & $x<x_{\rm h}$ (dynamical) & $x>x_{\rm h}$ (static) \\ \hline
Even & Eq.~(\ref{ansatz00-2}) with $M=-M_+(>0)$ & Eq.~(\ref{ansatz00-2}) with $M=M_+(<0)$ \\ \cline{1-3}
Odd & Eq.~(\ref{ansatz00-2-interior}) with ${\bar M}=M_+(<0)$ & Eq.~(\ref{ansatz00-2}) with $M=M_+(<0)$ \\ \cline{1-3}
Even & Eq.~(\ref{ansatz00-2-interior}) with ${\bar M}=-{\bar M}_+(<0)$ & Eq.~(\ref{ansatz00-2-interior}) with ${\bar M}={\bar M}_+(>0)$ \\ \cline{1-3}
Odd & Eq.~(\ref{ansatz00-2}) with $M={\bar M}_+(>0)$ & Eq.~(\ref{ansatz00-2-interior}) with ${\bar M}={\bar M}_+(>0)$ \\ \hline
\end{tabular}
}
\end{center}
\end{table*}

\section{Black hole and black bounce for $\chi\in(\chi_0,0)$}
\label{sec:main}

Proposition~\ref{Pro:Gamboa-extension1} shows that a variety of regular extensions beyond $x=x_{\rm h}$ are possible for the Gamboa solution given by Eq.~(\ref{ansatz00-2}) or (\ref{ansatz00-2-interior}) even with $\chi_+\ne \chi_-$ and/or different values of $M$ or ${\bar M}$.
In this section, we study the global structure of the maximally extended Gamboa solution given by Eq.~(\ref{ansatz00-2}) with $M<0$ and $h_1>0$ for $\chi\in(\chi_0,0)$ or, equivalently, $\beta\in(2/(n-3),\infty)$.
In this case, the spacetime is static and asymptotically S-T as $r\to \infty$ and $r=r_{\rm h}$ corresponds to a nondegenerate Killing horizon.
For simplicity, we consider the extension such that the value of $\chi$ is unchanged in the extended region (and hence $\beta$ and $h_1$ are as well).
Although the Gamboa solution can be attached to the Minkowski spacetime at $x=x_{\rm h}$, we assume here that the extended region is nonvacuum and study such an unexpected case in a separate paper~\cite{Maeda:2025ghc}.
As we will show, the solution describes either an asymptotically topological S-T globally regular black bounce or a black hole, depending on the parameters.


We clarify the maximally extended Gamboa solution given by Eq.~(\ref{ansatz00-2}) with $M=M_+<0$ and $h_1>0$ for $\chi\in(\chi_0,0)$, with which $\Pi_1>0$ and $\alpha>0$ are satisfied.
By Lemma~\ref{lm:Gamboa-quasi-global}, the solution is mapped onto the domain $x> x_{\rm h}$ in the single-null coordinates (\ref{metric-Buchdahl-v}).
Equation~(\ref{trans1}) with $\Omega/\Pi_1=-(2+\beta)/(M\Pi_1^2)$ gives
\begin{align}
&\frac{\D r}{\D x}=-\frac{M\Pi_1}{2+\beta}\Pi^{(1+\beta)/(2+\beta)},\label{r'1}
\end{align} 
which shows that $r(x)$ is a monotonically increasing function in this static domain.
According to the results in Sec.~\ref{sec:infinity}, the asymptotically S-T region $r\to \infty$ corresponding to $x\to \infty$ is null infinity and causally null in the Penrose diagram.
Under the assumption that $\beta$ is the same on both sides of the horizon, the continuity of $r(x)$ at $x=x_{\rm h}$ requires that $h_1$ (and hence $\Pi_1$ as well) is also the same on both sides of the horizon.

For $\chi\in(\chi_0,0)$, the matter field of the Gamboa solution in the form of Eq.~(\ref{ansatz00-2}) with $M\ne 0$ and $h_1>0$ violates all the standard energy conditions by Eqs.~(\ref{NEC-I})--(\ref{SEC-I}) for $M<0$ and by Eqs.~(\ref{NEC-I-interior})--(\ref{SEC-I-interior}) for $M>0$.
On the other hand, in the form of Eq.~(\ref{ansatz00-2-interior}), it satisfies all the standard energy conditions for ${\bar M}>0$ by Eqs.~(\ref{NEC-I})--(\ref{SEC-I}), while it satisfies the NEC and SEC but violates the WEC and DEC for ${\bar M}<0$ by Eqs.~(\ref{NEC-I-interior})--(\ref{SEC-I-interior}).
Those results combined with Lemma~\ref{lm:Gamboa-quasi-global} are summarized in Table~\ref{table:domain-ECs}.
\begin{table*}[htb]
\begin{center}
\caption{Domains of the Gamboa solution in the quasiglobal coordinates (\ref{metric-Buchdahl}) and the single-null coordinates (\ref{metric-Buchdahl-v}) for $\chi\in(\chi_0,0)$. The energy conditions satisfied in the domain are shown inside the square brackets.}
\label{table:domain-ECs}
\scalebox{1.0}{
\begin{tabular}{|c|c|c|c|c|}\hline
Form & $x<x_{\rm h}$ (dynamical) & $x>x_{\rm h}$ (static) \\ \hline
Eq.~(\ref{ansatz00-2}) ($r>r_{\rm h}$) & $M>0$ [None] & $M<0$ [None] \\ \cline{1-3}
Eq.~(\ref{ansatz00-2-interior}) ($r<r_{\rm h}$) & ${\bar M}<0$ [NEC \& SEC] & ${\bar M}>0$ [All] \\ \hline
\end{tabular}
}
\end{center}
\end{table*}

\begin{figure}[htbp]
\includegraphics[width=0.35\textwidth]{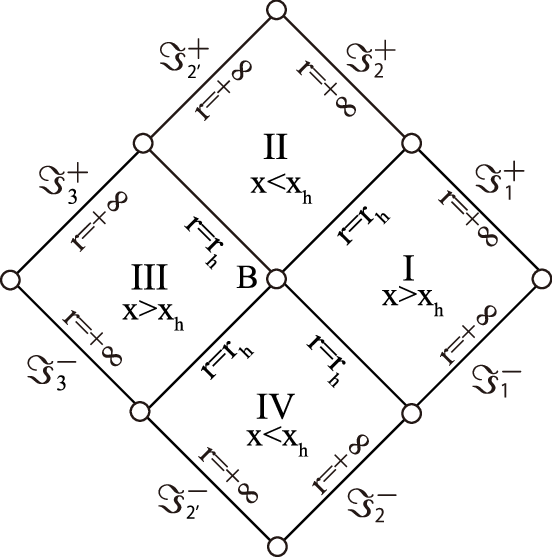}
\caption{
\label{Fig:NullBlackBounce-k=0} A black-bounce spacetime as a maximal extension of the asymptotically S-T Gamboa solution (\ref{ansatz00-2}) for $\chi=\chi_+\in(\chi_0,0)$ with $M=M_+(<0)$ and $h_1>0$ beyond the Killing horizon $x=x_{\rm h}$ under the assumption $\chi=\chi_+$ in the extended domain $x<x_{\rm h}$.
The extended dynamical domain $x<x_{\rm h}$ is described by the Gamboa solution in the form of Eq.~(\ref{ansatz00-2}) with $M=M_->0$.
$\Im^{+(-)}$ stands for a future (past) null infinity and the subscripts distinguish among different null infinities.
}
\end{figure}

Suppose that the extended domain $x<x_{\rm h}$, which is dynamical as $H<0$ holds, is described by the Gamboa solution in the form of Eq.~(\ref{ansatz00-2}) with $M=M_->0$.
Equation~(\ref{r'1}) with $M>0$ shows that $r(x)$ is a monotonically decreasing function in this dynamical domain.
According to the results in Sec.~\ref{sec:infinity}, the asymptotically S-T region $r\to \infty$ corresponding to $x\to -\infty$ is null infinity and causally null in the Penrose diagram.
As a result, the maximally extended spacetime given in the domain $-\infty<x<\infty$ in the coordinate system (\ref{metric-Buchdahl-v}) describes a globally regular black bounce for any $\chi\in(\chi_0,0)$ or, equivalently, $\beta\in[0,\infty)$, for which the Penrose diagram is drawn in Fig.~\ref{Fig:NullBlackBounce-k=0}.
The bounce hypersurface is null and coincides with the Killing horizon $r=r_{\rm h}$ which is also an event horizon and a wormhole throat. 
As shown in Table~\ref{table:C-infty}, the metric in the coordinate system (\ref{metric-Buchdahl-v}) is $C^\infty$ only for an even $\beta$ and $M_-=-M_+$.
All the standard energy conditions are violated everywhere except on the Killing horizon $x=x_{\rm h}$.

\begin{figure}[htbp]
\includegraphics[width=0.35\textwidth]{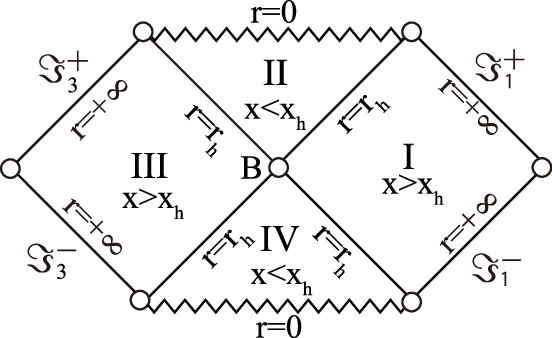}
\caption{
\label{Fig:BlackHole-k=0} A black-hole spacetime as a maximal extension of the asymptotically S-T Gamboa solution (\ref{ansatz00-2}) for $\chi=\chi_+\in(\chi_0,0)$ with $M=M_+(<0)$ and $h_1>0$ beyond the Killing horizon $x=x_{\rm h}$ under the assumption that $\chi=\chi_+$ in the extended domain $x<x_{\rm h}$.
The extended dynamical domain $x<x_{\rm h}$ is described by the Gamboa solution in the form of Eq.~(\ref{ansatz00-2-interior}) with ${\bar M}={\bar M}_-<0$.
A zigzag line indicates a curvature singularity located at $r=0$ corresponding to $x=x_{\rm s}(<x_{\rm h})$.
}
\end{figure}

Next, suppose that the extended dynamical domain $x<x_{\rm h}$ is described by the Gamboa solution in the form of Eq.~(\ref{ansatz00-2-interior}) with ${\bar M}={\bar M}_-<0$.
Equation~(\ref{trans1-interior}) with $\Omega=-(2+\beta)/({\bar M}\Pi_1)$ gives
\begin{align}
&\frac{\D r}{\D x}=-\frac{{\bar M}\Pi_1}{2+\beta}(-\Pi)^{(1+\beta)/(2+\beta)}>0,
\end{align} 
which shows that $r(x)$ is a monotonically increasing function in this dynamical domain.
Therefore, there is a curvature singularity at $x=x_{\rm s}(<x_{\rm h})$ determined by $r(x_{\rm s})=0$.
According to the results in Sec.~\ref{sec:infinity}, this singularity is spacelike in the Penrose diagram.
As a result, the maximally extended Gamboa spacetime given in the domain $x_{\rm s}<x<\infty$ in the coordinate system (\ref{metric-Buchdahl-v}) describes a black hole with a nondegenerate Killing horizon at $x=x_{\rm h}$ and a curvature singularity at $x=x_{\rm s}$ for any $\beta\in[0,\infty)$, for which Penrose diagram is drawn in Fig.~\ref{Fig:BlackHole-k=0}.
As shown in Table~\ref{table:C-infty}, the metric in the coordinate system (\ref{metric-Buchdahl-v}) is $C^\infty$ only for odd $\beta$ and ${\bar M}_-=M_+$.
While all the standard energy conditions are violated in the domain $x\ge x_{\rm h}$, only the NEC and the SEC are satisfied in the dynamical domain $x_{\rm s}<x<x_{\rm h}$.
The results are summarized in Table~\ref{table:structure1}.
\begin{table*}[htb]
\begin{center}
\caption{Maximal extensions of the asymptotically S-T Gamboa solution (\ref{ansatz00-2}) for $\chi=\chi_+\in(\chi_0,0)$ with $M=M_+(<0)$ and $h_1>0$ beyond the Killing horizon $x=x_{\rm h}$ under the assumption that $\chi=\chi_+$ in the extended domain $x<x_{\rm h}$. The energy conditions satisfied in the domain are shown inside the square brackets.}
\label{table:structure1}
\scalebox{0.95}{
\begin{tabular}{|c|c|c|c|c|}\hline
& $x<x_{\rm h}$ (dynamical) & $x>x_{\rm h}$ (static) & Diagram & $C^\infty$ horizon \\ \hline
Black bounce & Eq.~(\ref{ansatz00-2}) with $M=M_-(>0)$ [None] & Eq.~(\ref{ansatz00-2}) with $M=M_+(<0)$ [None] & Fig.~\ref{Fig:NullBlackBounce-k=0} & Even $\beta$, $M_-=-M_+$ \\ \cline{1-5}
Black hole & Eq.~(\ref{ansatz00-2-interior}) with ${\bar M}={\bar M}_-(<0)$ [NEC \& SEC] & Eq.~(\ref{ansatz00-2}) with $M=M_+(<0)$ [None] & Fig.~\ref{Fig:BlackHole-k=0} & Odd $\beta$, ${\bar M}_-=M_+$ \\ \hline
\end{tabular}
}
\end{center}
\end{table*}

In fact, using Proposition~\ref{Pro:Gamboa-extension1}, regions I--IV in Figs.~\ref{Fig:NullBlackBounce-k=0} and \ref{Fig:BlackHole-k=0} can be described by the Gamboa solution for different values of $\chi$ unless $\chi\in(\chi_0,0)$ is satisfied. (See also Proposition~6 in Ref.~\cite{addendum}.)
Even in such configurations, the bifurcation $(n-2)$-surface, denoted by B, is regular, as shown in Appendix~\ref{app:bifurcation}.
A physical explanation of this counterintuitive regularity was provided in Ref.~\cite{Maeda:2024tpl} based on the fact that the matter field in the dynamical region can be interpreted as a spacelike (tachyonic) perfect fluid.
A fluid element of such a fluid in regions II and IV moves in a spacelike direction and does not cross either a Killing horizon or a bifurcation surface.
Because orbits of the fluid elements do not have a past or future endpoint at B, the bifurcation $(n-2)$-surface B can be regular.

\section{Summary}

In this paper, we have investigated the Gamboa solution given in the form of Eq.~(\ref{ansatz00-2}) or (\ref{ansatz00-2-interior}) in general relativity, which is an exact two-parameter family of plane symmetric solutions admitting a hypersurface-orthogonal Killing vector $\xi^\mu\partial_\mu=\partial_T$ with a perfect fluid obeying an equation of state $p=\chi\rho$ in $n(\ge 4)$ dimensions~\cite{GamboaSaravi}.
Proposition~\ref{Pro:Gamboa-extension1} is one of our main results that identifies all possible regular attachments of two Gamboa solutions for $\chi\in[-1/3,0)$ or, equivalently, $\beta\in[0,\infty)$ at the Killing horizon $x=x_{\rm h}$ in the regular coordinate system (\ref{metric-Buchdahl-v}).
We have also derived the conditions for the $C^\infty$ attachment at the horizon.

The other main result is that we have clarified the maximally extended spacetimes of the asymptotically topological S-T Gamboa solution given by Eq.~(\ref{ansatz00-2}) with $M=M_+<0$ and $h_1>0$ for $\chi\in(\chi_0,0)$ or, equivalently, $\beta\in(2/(n-3),\infty)$ under the assumption that the value of $\chi$ is unchanged in the extended dynamical region.
This result is summarized in Table~\ref{table:structure1}.
In more detail, we have shown the following.
\begin{enumerate}
\item For any value of $\chi\in(\chi_0,0)$, the maximally extended spacetime describes (i) a globally regular and asymptotically topological S-T black bounce with a nondegenerate Killing horizon which coincides with a bounce null hypersurface, or (ii) an asymptotically topological S-T black hole with a nondegenerate Killing horizon and a spacelike curvature singularity inside the horizon.
\item In the black-bounce case, the metric in the single-null coordinates (\ref{metric-Buchdahl-v}) is $C^\infty$ only for $\chi=-1/(1+2N)$ with odd $N$ satisfying $N>(n-1)/(n-3)$, and if the parameter in the extended region is given by $M=-M_+(>0)$.
\item In the black-hole case, the metric in the single-null coordinates (\ref{metric-Buchdahl-v}) is $C^\infty$ only for an odd $\beta$ or, equivalently, $\chi=-1/(1+2N)$ with even $N$ satisfying $N>(n-1)/(n-3)$, and if the parameter in the extended region is given by ${\bar M}=M_+(<0)$.
\end{enumerate}

Because the NEC is violated away from the Killing horizon, our black-bounce spacetime is expected to be dynamically unstable.
Nevertheless, it is intriguing that such a configuration is possible as a plane symmetric solution because the vacuum solution does not describe a black hole unless a negative cosmological constant is introduced.
Of course, it would be even more interesting if such a configuration were possible as an asymptotically flat spherically symmetric solution with a perfect fluid.
This problem is left for future investigation.

\acknowledgments

H.M. is very grateful to Max-Planck-Institut f\"ur Gravitationsphysik (Albert-Einstein-Institut) and Centro de Estudios Cient\'{\i}ficos, where a large part of this work was carried out, for their hospitality and support. 
This work has been partially supported by ANID FONDECYT Grants No.~1220862 and No.~1241835.

\appendix

\section{General static solution with $p=\chi \rho$ and a Ricci-flat base manifold.}
\label{app:General-sol}

In this appendix, we derive the most general $n(\ge 4)$-dimensional static solution with an $(n-2)$-dimensional Ricci-flat base manifold $K^{n-2}$ in general relativity with a perfect fluid obeying a linear equation of state $p=\chi\rho$ given by Eq.~\eqref{EFE-0}.
In this system, without loss of generality, one may adopt the comoving coordinates with the radial coordinate $r$ as the areal radius such that
\begin{align}
\label{ansatz0}
\begin{aligned} 
&\D s^2=-A(r)\D T^2+B(r)\D r^2+r^2\gamma_{ij}(z)\D z^i\D z^j,\\
&u^\mu\frac{\partial}{\partial x^\mu}=\frac{1}{\sqrt{A}}\frac{\partial}{\partial T},
\end{aligned} 
\end{align} 
where $i,j = 2, 3,\cdots, n-1$ and $\gamma _{ij}$ is the metric on $K^{n-2}$.
This problem was addressed in Ref.~\cite{GamboaSaravi} using the comoving coordinates but with a nonareal radial coordinate and assuming the $(n-2)$-dimensional flat base manifold $\mathbb{R}^{n-2}$. The generalization including a Ricci-flat base manifold is straightforward since the Einstein equations only require the Ricci tensor of $K^{n-2}$, which is null in this case. 
However, the symmetries and global geometric properties of the spacetime can drastically differ after the base manifold is changed. 

In the coordinate system (\ref{ansatz0}), the Einstein equations \eqref{EFE-0} are written as
\begin{align}
\kappa_n\rho=&\frac{n-2}{2r^2B}\left\{\frac{rB'}{B}-(n-3)\right\}, \label{EFE00} \\
\kappa_n p=&\frac{n-2}{2r^2B}\left\{\frac{rA'}{A}+(n-3)\right\}, \label{EFE11} \\
\kappa_n p=&\frac{1}{2r^2B}\biggl[-\frac{rB'}{2B}\left\{\frac{rA'}{A}+2(n-3)\right\}+\frac{r^2A''}{A} \nonumber \\
&+(n-3)\frac{rA'}{A}-\frac{r^2{A'}^2}{2A^2}+(n-3)(n-4)\biggl].\label{EFEsub}
\end{align}
The energy-momentum conservation equations $\nabla_\nu T^{\mu\nu}=0$ give
\begin{align}
p'=-\frac{A'}{2A}(\rho+p).\label{cons}
\end{align} 
Now we assume the linear equation of state $p=\chi\rho$.
The dust case ($\chi=0$) is excluded from our consideration because Eq.~(\ref{cons}) shows that $A$ is constant and that it contradicts to Eq.~(\ref{EFE11}) for $n\ge 4$, so there is no static dust solution.
In addition, Eq.~(\ref{cons}) shows that a perfect fluid with $\chi=-1$ is equivalent to a cosmological constant $\Lambda$. 
Hence, by Birkhoff's theorem with $\Lambda$, the general solution is identical to the topological Schwarzschild-Tangherlini-(anti-)de~Sitter solution.

With $p=\chi \rho$ and $\chi \neq 0$, Eq.~(\ref{cons}) is integrated, yielding
\begin{align}
\rho=c_0 A^{-(1+\chi)/(2\chi)},\label{rho-A}
\end{align} 
where $c_0$ is a constant.
Moreover, Eqs.~(\ref{EFE00}) and (\ref{EFE11}) leads to
\begin{align}
&\chi\frac{ rB'}{B}=\frac{rA'}{A}+(n-3)(\chi+1), \label{EFE-part1}
\end{align}
which is easily integrated as
\begin{align} \label{BA}
B=&B_0r^{(n-3)(\chi+1)/\chi}A^{1/\chi},
\end{align}
where $B_0$ is an integration constant.
In addition, Eq.~(\ref{EFE11}) provides the following expression for $\rho$:
\begin{align} \label{rho2}
\rho=&\frac{n-2}{2\chi \kappa_nB_0r^{[(n-1)\chi+(n-3)]/\chi}A^{1/\chi}}\left\{\frac{rA'}{A}+(n-3)\right\}.
\end{align}

Introducing a new variable $a(r)$ such that
\begin{align} \label{AApend}
A(r)=r^{-(n-3)}a(r),
\end{align}
we obtain from Eqs. \eqref{BA} and \eqref{rho2} 
\begin{align}
B=&B_0r^{n-3}a^{1/\chi},\label{Bsol}\\
\rho=&\frac{(n-2)a'}{2\chi \kappa_nB_0r^{n-2}a^{(1+\chi)/\chi}}, \label{rho-alpha}
\end{align}
respectively. The comparison of Eqs.~\eqref{rho-A} and \eqref{rho-alpha} then yields the first-order differential equation
\begin{align} \label{Ap2}
\begin{aligned}
(n-2)a'=&2\chi \kappa_nc_0 B_0r^{(\zeta-2\chi)/(2\chi)}a^{(1+\chi)/(2\chi)},\\
\zeta:=&(3n-5)\chi+(n-3),
\end{aligned}
\end{align}
which can be directly integrated by quadrature. The general solution to Eq.~(\ref{Ap2}) is 
\begin{align}
&a(r)=\left\{
\begin{array}{ll}
h^{-2\chi/(1-\chi)} & [\chi\ne 0,1],\\
\exp(A_0+A_1r^{2(n-2)}) & [\chi=1],
\end{array}
\right.
\end{align}
where $h(r)$ is defined by 
\begin{align} \label{hApend}
\begin{aligned}
h(r):=&\left\{
\begin{array}{ll}
A_0+A_1r^{\zeta/(2\chi)} & [\chi\ne \chi_0],\\
A_0+A_1\ln|r| & [\chi=\chi_0],
\end{array}
\right.\\
\chi_0:=&-\frac{n-3}{3n-5}.
\end{aligned}
\end{align}
Here $A_0$ is an integration constant and the constant $A_1$ is given by 
\begin{align} \label{A1app}
& A_1=\left\{
\begin{array}{ll}
-2\kappa_nc_0B_0\chi(1-\chi)/[(n-2)\zeta] & [\chi\ne 0,1,\chi_0],\\
-4\kappa_nc_0B_0/(3n-5) & [\chi=\chi_0],\\
\kappa_nc_0B_0/(n - 2)^2 & [\chi=1].
\end{array}
\right.
\end{align}

Although four integration constants appear in the general solution, Eq.~\eqref{A1app} reduces this number to three. 
In addition, the field equations given by Eqs.~\eqref{EFE00}--\eqref{cons} are clearly invariant under a constant rescaling of the lapse function $A(r)$, which is equivalent to a rescaling of the time coordinate $T$. 
Thus, another constant can be fixed without loss of generality. Therefore, the general solution is characterized by two parameters. 
One of these two parameters is associated with the conserved charge, the mass corresponding to the time translation invariance of a static configuration, and the other is associated with the presence of a matter field. 

Lastly, we briefly comment on the case $\chi=1$. By means of a rescaling of $T$ and a redefinition of the integration constants, a suitable form of the general solution for $\chi=1$ is given by
\begin{align}
\label{sol-chi=1-best}
\begin{aligned}
\D s^2=&-{ B}_0^{-1}r^{-(n-3)}\exp(A_1r^{2(n-2)})\D T^2\\
&+{ B}_0r^{n-3}\exp(A_1r^{2(n-2)})\D r^2+r^2\gamma_{ij}\D z^i\D z^j,\\
\rho&=p=\frac{(n-2)^2A_1r^{n-3}}{\kappa_n{ B}_0\exp(A_1r^{2(n-2)})}.
\end{aligned}
\end{align} 
The solution (\ref{sol-chi=1-best}) with $A_1=0$ is identical to the topological Schwarzschild-Tangherlini vacuum solution.
In four dimensions ($n=4$), and when the base manifold is chosen as $\mathbb{R}^2$, this solution reproduces the static plane symmetric one obtained in Ref.~\cite{Tabensky-Taub1973}.

\section{Regularity of the bifurcation $(n-2)$-surface}
\label{app:bifurcation}

In this appendix, we show the regularity of the bifurcation $(n-2)$-surface, denoted by B in Figs.~\ref{Fig:NullBlackBounce-k=0} and \ref{Fig:BlackHole-k=0}.
For static and plane symmetric solutions with $\chi\in[-1/3,0)$, the asymptotic behaviors of the metric functions near a nondegenerate Killing horizon $x=x_{\rm h}$ in the domain $x> x_{\rm h}$ are given by Eq.~(\ref{Asymp-sol}).
We now write the metric near $x=x_{\rm h}$ in the domain $x> x_{\rm h}$ as
\begin{align}
\label{expand+}
\begin{aligned}
&H(x)\simeq H_1\Delta+H_{3+\beta}^+\Delta^{3+\beta},\\
&r(x)\simeq r_{\rm h}+r_{2+\beta}^+\Delta^{2+\beta},
\end{aligned} 
\end{align} 
and in the domain $x< x_{\rm h}$ as
\begin{align}
\label{expand-}
\begin{aligned}
&H(x)\simeq H_1\Delta+H_{3+\beta}^-(-\Delta)^{3+\beta},\\
&r(x)\simeq r_{\rm h}+r_{2+\beta}^-(-\Delta)^{2+\beta},
\end{aligned} 
\end{align} 
where $\Delta:=x-x_{\rm h}$ and $\beta\in[0,\infty)$ is defined by Eq.~(\ref{def-beta}).
Here $H_1$ and $r_{\rm h}$ are nonzero in particular.
Although the values of the coefficients of the higher-order terms are, in general, different, the Killing horizon $x=x_{\rm h}$ is regular as the metric in the single-null coordinates (\ref{metric-Buchdahl}) is at least $C^{1,1}$ there.

Using the tortoise coordinate $x_*:=\int H(x)^{-1}\D x$, we define the null Kruskal-Szekeres coordinates $U$ and $V$ by
\begin{align}
U:=\mp e^{-H_1u/2},\qquad V:=e^{H_1v/2},\label{def-UV}
\end{align} 
where $u:=t-x_*$ and $v:=t+x_*$.
For a nondegenerate outer Killing horizon characterized by $H_1>0$, $x=x_{\rm h}$ corresponds to $x_*\to -\infty$ and $UV= 0$.
The upper (lower) sign in the definition of $U$ in Eq.~(\ref{def-UV}) corresponds to $x>(<)x_{\rm h}$.
The metric (\ref{metric-Buchdahl}) in the quasiglobal coordinates is then written in the null Kruskal-Szekeres coordinates as
\begin{align}
\label{Kruskal}
\begin{aligned}
&\D s^2=2g_{UV}\D U\D V+r(x(U,V))^2\D l_{n-2}^2,\\
&g_{UV}=\frac{2H(x(U,V))}{H_1^2UV}.
\end{aligned} 
\end{align} 
With both Eq.~(\ref{expand+}) and Eq.~(\ref{expand-}), the asymptotic behaviors of the metric functions near a Killing horizon $x_*\to -\infty$ ($UV=0$) are given by 
\begin{align}
\label{expansion-UV}
\begin{aligned}
g_{UV}\simeq& \frac{2}{H_1^2}\biggl(-1+O(U^3V^3)\biggl),\\
r^2\simeq& r_{\rm h}^2+O(U^2V^2).
\end{aligned} 
\end{align} 
Equation~(\ref{expansion-UV}) shows that the metric is at least $C^{1,1}$ and hence regular also at a bifurcation $(n-2)$-surface given by $U=V=0$.

\section{Matter field on the Killing horizon for $\chi=-1/3$}
\label{app:geodesic}

In this appendix, we derive an explicit form of the matter field confined on the Killing horizon $x=x_{\rm h}$ in the Gamboa solution for $\chi=-1/3$.
For $\chi=-1/3$, the matter field at $x=x_{\rm h}$ is a null dust fluid by Proposition~2 and Corollary~1 in Ref.~\cite{Maeda:2021ukk}.
Using Eq.~(\ref{asymp-H1}) with $\chi=-1/3$ ($\beta=0$), we obtain its energy-momentum tensor as
\begin{align}
&T_{\mu\nu}|_{x=x_{\rm h}}=\sigma k_\mu k_\nu,\qquad k^\mu\frac{\partial}{\partial x^\mu}={\cal K}\frac{\partial}{\partial v},\label{typeII-null}
\end{align}
where $\sigma$ is the energy density of the null dust and $k^\mu$ is a null vector satisfying $k_\mu k^\mu=0$ and 
\begin{align}
\sigma{\cal K}^2=&-\frac{(n-2)r''}{\kappa_n r}\biggl|_{x=x_{\rm h}}=-\frac{4(n-2)M^2}{\kappa_n r_{\rm h}^4}.
\end{align}
Since $\sigma<0$ is satisfied, the null dust confined on the Killing horizon violates all the standard energy conditions.
(See Sec.~4.2 in Ref.~\cite{Maeda:2018hqu}.)
The tangent vector $k^\mu$ is arbitrary to multiply by a scalar function because it is null.
As a natural choice, we choose $k^\mu$ as a null generator of the Killing horizon.

Consider an affinely parametrized future-directed radial null geodesic $\gamma$ in the spacetime described by the metric in the single-null coordinates (\ref{metric-Buchdahl-v}).
The orbit of $\gamma$ is given by $x^\mu=(v(\lambda),x(\lambda),0,\cdots,0)$ with its tangent vector $k^\mu=({\dot v},{\dot x},0,\cdots,0)$, where $\lambda$ is an affine parameter along $\gamma$ and the dots denote differentiation with respect to $\lambda$.
Then null geodesic equations $k^\nu\nabla_\nu k^\mu=0$ for $\gamma$ are written as
\begin{align}
0=&{\ddot v}+\frac12H'{\dot v}^2,\\
0=&{\ddot x}+\frac12HH'{\dot v}^2-H'{\dot v}{\dot x},
\end{align} 
which admit the following solution:
\begin{align}
v(\lambda)=&r_{\rm h}^{n-3}\ln|\lambda-\lambda_0|+v_0,\qquad x(\lambda)=x_{\rm h},\label{sol-geodesic}
\end{align} 
where we have used $H(x_{\rm h})=0$ and $H'(x_{\rm h})=2/r_{\rm h}^{n-3}$ given by Eq.~(\ref{asymp-H1}) and $v_0$ and $\lambda_0$ are integration constants.

The solution (\ref{sol-geodesic}) is a null generator of the Killing horizon $x=x_{\rm h}$ and $\lambda\to \lambda_0$ corresponds to a bifurcation $(n-2)$-surface $v\to -\infty$ on which the Killing vector generating staticity vanishes.
Identifying the solution (\ref{sol-geodesic}) as the orbit of a null dust fluid on the Killing horizon given by Eq.~(\ref{typeII-null}), we obtain $\sigma$ and ${\cal K}$ as 
\begin{align}
\sigma=-\frac{4(n-2)M^2}{\kappa_n r_{\rm h}^{2(n-1)}}(\lambda-\lambda_0)^2,\qquad {\cal K}=\frac{r_{\rm h}^{n-3}}{\lambda-\lambda_0}.
\end{align}
The energy density $\sigma$ converges to zero in the limit to a bifurcation $(n-2)$-surface $\lambda\to \lambda_0$.


\end{document}